\def\ZPC#1#2#3{{\sl Z.~Phys.} {\bf C#1}~(#3) #2}

\def\PRL#1#2#3{{\sl Phys. Rev. Lett.} {\bf #1}~(#3) #2}

\def\PLB#1#2#3{{\sl Phys. Lett.} {\bf B#1}~(#3) #2}

\def\NPB#1#2#3{{\sl Nucl. Phys.} {\bf B#1}~(#3) #2}

\def\beq{\begin{equation}}
\def\eeq{\end{equation}}
\def\bea{\begin{eqnarray}}
\def\eea{\end{eqnarray}}
\def\bq{\begin{quote}}
\def\eq{\end{quote}}

\def\gappeq{\mathrel{\rlap {\raise.5ex\hbox{$>$}}
{\lower.5ex\hbox{$\sim$}}}}

\def\lappeq{\mathrel{\rlap{\raise.5ex\hbox{$<$}}
{\lower.5ex\hbox{$\sim$}}}}

\def\bbz{fa Z \kern-8.9pt Z}
\documentstyle [12pt,epsfig]{article}
  \newcommand{\ccaption}[2]{
    \begin{center}
    \parbox{0.85\textwidth}{
      \caption[#1]{\small{\it{#2}}}
      }
    \end{center}
    }
\begin{document}
\thispagestyle{empty}
\vspace*{-1cm}
\begin{flushright}
{CERN-TH/97-109} \\
{TIFR-TH-97-25} \\
{hep-ph/9705416} \\
\end{flushright}
\vspace{1cm}
\begin{center}
{\large {\bf Pursuing the Strange Stop Interpretation of the HERA
Large-Q$^2$ Data}} \\
\vspace{.2cm}                                                  
\end{center}

\begin{center}
\vspace{.3cm}
{\bf J. Ellis,                                                                
S. Lola} and                                                     
{\bf K. Sridhar}\footnote{Permanent Address:
Theory Group, Tata Institute of Fundamental Research,
Homi Bhabha Road, Bombay~400~005, India.} 
\\                          
\vspace{.5cm}
{ Theory Division, CERN, CH-1211, Gen\`eve 23, Switzerland} \\
\end{center}                              

\vspace{2cm}

\begin{abstract}
We explore the possible interpretation of the large-$Q^2$ anomaly
reported by the H1 and ZEUS collaborations in terms of stop squark
production off a strange quark in the proton via an $R$-violating
interaction. This ``strange stop" interpretation is constrained by LEP
measurements of the $Z \rightarrow e^+ e^-$ decay rate 
in addition to constraints from the electroweak $\rho$ parameter and
CDF and D0 searches for first-generation leptoquarks. We
investigate the interplay between these constraints, taking full
account of stop mixing effects. We find that if $m_{\tilde t} \le 200$
GeV only relatively small domains of the chargino and neutralino
parameters are consistent with these constraints,
and explore the extent to which this scenario may be probed further
by searches for contact interactions at LEP~2 and experiments with
$e^-$ and polarized beams at HERA.
\end{abstract}
\vfill 
\begin{flushleft}
CERN-TH/97-109\\
May 1997\\
\end{flushleft}

\newpage

The surprisingly large number of high-$x$ and -$Q^2$ deep-inelastic 
$e^+ p$ scattering events at HERA, 
reported recently by the H1 and ZEUS collaborations~\cite{H1,ZEUS}, is
currently exciting considerable theoretical 
speculation~\cite{interprets,AEGLM,KRSZ,KK}.
It may well be that the excess reported is just  a statistical fluctuation,
and this may become apparent from the additional data being obtained
during the current year's run. However, unless and until this happens, it
is the responsibility of theorists to pursue different possible 
interpretations
of the excess, assuming it to be real, evaluate and combine the
constraints imposed by other data, and propose additional signatures
that may serve to discriminate between the surviving rival interpretations.
Effective contact interactions from some energy scale beyond that of the 
Standard Model do not seem to match well the combined $Q^2$ distribution
reported by the H1 and ZEUS collaborations. Interest has been piqued
by the tendency of the H1 data to peak in an $x$ range
corresponding to the production of a narrow resonance with a mass around
200 GeV~\cite{H1} and leptoquark quantum numbers, 
which is apparently not contradicted by 
the ZEUS data~\cite{AEGLM,Wolf}.

If one chooses to speculate along this line, upper limits on the
production cross section at the Fermilab Tevatron collider~\cite{D0,CDF}
strongly suggest that any such leptoquark state must have spin 0, 
and hint that its branching ratio 
${\cal B}(e^+ q)$ into the observed final state
should be significantly less than unity~\cite{AEGLM}. 
The latter is difficult, though
not impossible, to arrange in a model without supersymmetry, but is
natural if the ``leptoquark" is actually a squark produced via an 
interaction that violates $R$ parity~\cite{pioneers1,pioneers2,pioneers3}: 
$\lambda'_{ijk} L_i Q_j D^c_k$, where $L_i, Q_j$ denote left-handed lepton
and quark doublets, and $D^c_k$ denotes left-handed charge-$1/3$ antiquark
singlets. In such a supersymmetric scenario, there
may be significant competition between the branching ratio
$\cal B$ for $R$-violating
decays
via the $\lambda'_{1jk}$ production coupling and $R$-conserving
decay modes such as $\chi q$ followed by $\chi \rightarrow \ell 
{\bar q} q$ decay~\cite{AEGLM}. Within this general supersymmetric
framework, three possible interpretations seem to emerge: $e^+_R d_R 
\rightarrow {\tilde c}_L, e^+_R d_R \rightarrow {\tilde t}$ and
$e^+_R s_R \rightarrow {\tilde t}$.

In this paper, we focus on the latter ``strange stop" interpretation,
which requires a relatively large $R$-violating coupling
$\lambda'_{132} \sim 0.3/(\hbox{cos} \phi \sqrt{\cal B})$, where $\phi$ is 
an angle describing mixing
between the ${\tilde t}_{L,R}$~\footnote{This must be significant, because of
the electroweak $\rho$ parameter~\cite{AEGLM}.}, and possibly
fine tuning of the
chargino spectrum so as to have a significant, but not too 
dominant, branching ratio into $\chi^+ b$~\cite{AEGLM}. As was
pointed out in~\cite{AEGLM}, this scenario must
contend with LEP~2 constraints on virtual $\tilde t$ exchanges
from the OPAL collaboration~\cite{OPAL}, as well as with upper limits from
precision electroweak physics on the possible contribution of the
${\tilde t}, {\tilde b}$ sector to the $\rho$ parameter. In addition, as
was pointed out in~\cite{Sridhar}, the measurement of the
$Z \rightarrow e^+ e^-$ decay rate provides an upper bound on the possible
magnitude of any $\lambda'_{13k}$ coupling, which is of
particular interest because it is independent of
the stop mixing angle and decay branching ratio. As we
shall see,
new data strengthen significantly the constraint found in the previous
analysis, 
providing stimulus for the re-evaluation in this paper of all the
phenomenological constraints on the ``strange stop" interpretation of the
HERA data, taking into account ${\tilde t}_{L,R}$ mixing effects~\footnote{We 
also comment, where appropriate, on implications for the ``down stop"
interpretation.}. 
We do not comment on possible
stringent limits on $R$-parity violating interactions from
cosmological considerations~\cite{CDEO}, as they can be avoided
in various schemes, such as baryogenesis at the electroweak scale
~\cite{Barel}.
We analyze the restrictions imposed by the $Z \rightarrow e^+ e^-$
and other constraints in the $\mu, M_2$ plane that characterizes
chargino and neutralino properties~\footnote{We
assume universality at the supersymmetric GUT scale for the input
soft supersymmetry-breaking gaugino masses $M_{1,2}$.}, 
and hence stop decay modes.
If the stop mass is 200 GeV or below, there is a restricted region
of this plane where $\lambda'_{132}$ is sufficiently small to satisfy
the $Z \rightarrow e^+ e^-$ constraint and ${\cal B}(e^+ s)$ is
sufficiently small to avoid the Tevatron production constraint. The
latter imposes no constraint if the stop mass exceeds 210 GeV, in which
case a significantly
larger region of the $\mu, M_2$ plane is allowed. We find that
searches for effective contact interactions at LEP~2 may have a limited
chance to probe further the ``strange stop" scenario, and also comment
on the possibilities for $e^-$ or polarized beams at HERA to cast
light on this scenario.

Any coupling of the form $\lambda^{\prime}_{i3k}$ would make
an extra contribution to the $Z$ partial width $\Gamma_{l_i} \equiv \Gamma
(Z \rightarrow 
l^+_i l^-_i)$ via loop diagrams
involving a $t$ and a $\tilde d_k$,
which is proportional to $m_t^2$ for large $m_t$. Since this may
become quite
large, precise determinations of the $\Gamma_{l_i}$ 
at LEP can provide sensitive constraints on the
possible magnitude of $R$ violation in the top sector~\cite{Sridhar}.
There are also contributions to $\delta \Gamma_l$ from diagrams
with stops in the loops, but
these are suppressed by a factor $m_i^2/m_t^2$ as compared to
the dominant contribution, where $i= d, s$ for the couplings
relevant to the interpretation of the HERA anomaly. These
diagrams have negligible numerical
impact on the bounds that we derive, so 
the effects of stop mixing on our bounds are also small. We note
that the masses of the $\tilde d_k$ could in principle be different
from that of the $\tilde t$ assumed to be produced at HERA. However,
the numerical values of the loop diagrams are not very sensitive to
$m_{\tilde d_k}$, which we assume for definiteness to equal $m_{\tilde
t}$.

We evaluate the top-${\tilde d_k}$ loops using the full expressions for
their contribution $\delta \Gamma_{l_i}$ given 
in Ref.~\cite{Sridhar}. It is clear that
the dominant contributions come from diagrams with
a top quark and a $\tilde d$ or $\tilde s$ squark in the
$\lambda'_{131}$ and $\lambda'_{132}$ cases, respectively.
The non-observation of
flavour-changing decays such as $\mu \rightarrow e \gamma$
imposes upper limits on the possibility that the lepton-number violation
induced by
$\lambda'$ couplings can be present simultaneously in both
$e (i=1)$ and $\mu (i=2)$ channels:
$\lambda'_{132} \lambda'_{232} < 0.015$ for $m_{\tilde t} \sim 200$
GeV~\cite{nomuande}.
This implies that these couplings cannot affect significantly both
$\Gamma_e$ and $\Gamma_{\mu}$~\footnote{This also means that ${\cal
B}(\mu^+ s) << {\cal B}(e^+ s)$ in the ``strange stop" scenario.}, so we
improve
the bounds on the $\lambda'_{13k}$ couplings presented in Ref.~\cite{Sridhar},
by considering the ratio $\Gamma_e/\Gamma_{\mu}$~\footnote{This 
is equivalent to considering the ratio $R_e/R_{\mu}$,
because the corrections the hadronic width $\Gamma_h$ from loops with
sleptons cancel when one takes the ratio.}. 
We have used the following experimental values for the 
leptonic widths \cite{lep}:
$\Gamma_e = 83.94 \pm 0.15$ MeV and
$\Gamma_{\mu} = 83.79 \pm 0.21$ MeV.
The resulting bounds on $\lambda'_{13k}$ are shown in
Fig.~1. We see that the bound on $\lambda^{\prime}$ 
varies from about 0.52 to about 0.6
as the squark mass changes from 180 to 220 GeV.
We emphasize that this bound is independent of stop mixing and
the ${\tilde t} \rightarrow e^+ q$ branching ratio $\cal B$,
making it a valuable constraint on
``strange stop" interpretation of the HERA anomaly.

\begin{figure}             
\centerline{\epsfig{figure=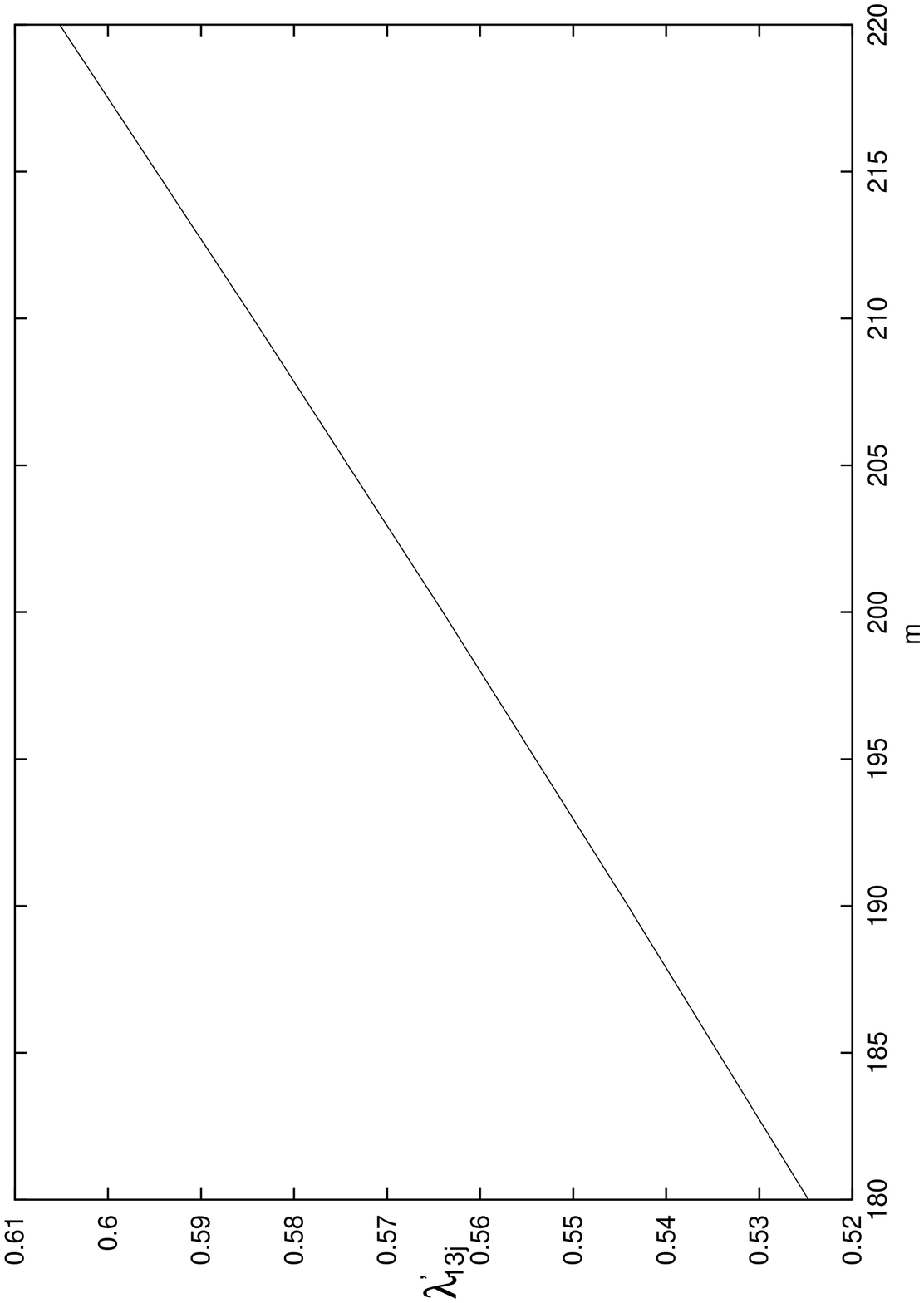,width=0.72\textwidth}}
\ccaption{}{ \label{fig:Zwidth}    
Upper bound on a generic $R$-violating coupling $\lambda'_{13k}$
obtained from $Z \rightarrow e^+ e^-, \mu^+ \mu^-$ decays, using the data 
in~\cite{lep} and the formulae in~\cite{Sridhar}.}
\end{figure}

The mass splitting in the stop-sbottom sector results in a 
contribution to the electroweak $\rho$ parameter, which is
constrained by
precision $Z$ measurements from LEP and the SLD, and by
$m_W/m_Z$ measurements from the Tevatron and elsewhere.
These and the Tevatron measurement of $m_t$ provide strong
upper limits on any contribution to $\rho$ from physics
beyond the Standard Model, in particular from the stop-sbottom
sector. In the absence of stop mixing \cite{AEGLM},
the $\tilde b_L$ mass is given by 
$m_{\tilde b_L} = \sqrt{m_{\tilde t_L^2} - {\rm cos} 2  
\beta  M_W^2 -m_t^2+m_b^2}$, and is about 100 to 130 GeV, which
would result in a
contribution $\Delta \rho$ larger than that allowed
by the data at the 95 \% C.L.~\cite{AEGLM}. We now
evaluate the amount of mixing that is required to avoid any
contradiction with the experimental upper limit on 
$ \Delta \rho$.

We neglect the bottom quark mass,
which also implies that we may neglect
mixing in the sbottom sector. The mixing in the stop
sector may be expressed in terms of an angle $\phi$ given by
\begin{equation}
{\rm tan} \phi = {m^2_{\tilde t_1} -m^2_{\tilde b_L}-m_t^2
-m_W^2{\rm cos}2\beta \over m_t m_{LR}} .
\end{equation}
where $m_{\tilde b_L}$ is the mass of
the physical left-handed sbottom, and we parametrize by $m_t
m_{LR}$ the off-diagonal element
in the stop mass-squared matrix where $m_{LR}$ is a combination
of the soft supersymmetry-breaking $A$ parameter and the Higgs
mixing parameter $\mu$. We denote by
$m_{\tilde t_{1,2}}$ the mass eigenstates that result from the mixing:
\begin{eqnarray}
m^2_{\tilde t_{1,2}} &=& {1\over 2}(m_{\tilde b_L}^2 +m_{\tilde t_R}^2
+2m_t^2+m_W^2{\rm cos}2\beta(1+{2\over 3}{\rm tan}^2\theta_W) \nonumber \\
&&\mp \lbrace\lbrack (m_{\tilde b_L}^2 -m_{\tilde t_R}^2
+m_W^2{\rm cos}2\beta(1-{2\over 3}{\rm tan}^2\theta_W) \rbrack^2
+4m_t^2 m_{LR}^2\rbrace^{1 \over 2}) ,
\end{eqnarray}
where $m_{\tilde t_R}^2$ is the soft supersymmetry-breaking contribution
to the right-handed stop mass. The contribution to $\Delta\rho$
coming from the $\tilde t - \tilde b$ mass-splitting is given
in this notation by \cite{inami, drees}
\begin{eqnarray}
\Delta\rho(\tilde t - \tilde b) &=& {3 \alpha \over 8 \pi m_W^2}
{\rm sin}^2\theta_W \biggl \lbrace {\rm cos}^2\phi f_s(m_{\tilde
t_1}^2, m_{\tilde b_1}^2) \nonumber \\
&& {\rm sin}^2\phi f_s(m_{\tilde t_2}^2, m_{\tilde b_1}^2) 
- {\rm cos}^2\phi {\rm sin}^2\phi f_s(m_{\tilde t_1}^2, m_{\tilde 
t_2}^2) \biggr\rbrace .
\end{eqnarray}
where
\begin{equation} 
f_s(m_1^2,m_2^2)={m_1^2m_2^2 \over m_1^2-m_2^2}{\rm ln}
{m_2^2 \over m_1^2} + {1 \over 2} (m_1^2+m_2^2)
\end{equation} 
We note that the $(t,b)$-loop contribution to $\rho$ is positive,
as is the range of $\rho$ indicated by the precision electroweak data.

We plot in Fig.~2 the ${\tilde t},{\tilde b}$ contribution to $\Delta\rho$
as a
function of the mixing angle $\phi$ for different choices of 
${\rm tan}\beta$ and $m_{LR}$ for $m_{\tilde t_1}=200$~GeV.
The parameter $m_{LR}$ cannot
be much larger than $m_{\tilde b_L}$, because of the
problem of false vacua \cite{frere}. As a rule of thumb, one may
impose $0 \lappeq m_{LR} \lappeq 3m_{\tilde b_L}$,
in which case $sin\phi < 0$. For the purposes
of this analysis we choose $m_{LR} = km_{\tilde b_L}$, with
$k=1,2$. 
The curves marked (a) and (b) in Fig.~2 are for
the case $k=1$ and ${\rm tan}\beta = 1,5$ respectively, whereas
those marked (c) and (d) are for $k=2$ and ${\rm tan}\beta =1,5$.
Curve (e) represents the 95 \% C.L. upper limit
after subtracting the Standard Model reference value for $m_H=100$ GeV,
as is appropriate for the MSSM.
We see that for $k=1$ and ${\rm tan}\beta=1$,
values of $\phi > -0.6$ are excluded for $m_H=100$ GeV.

\begin{figure}             
\centerline{\epsfig{figure=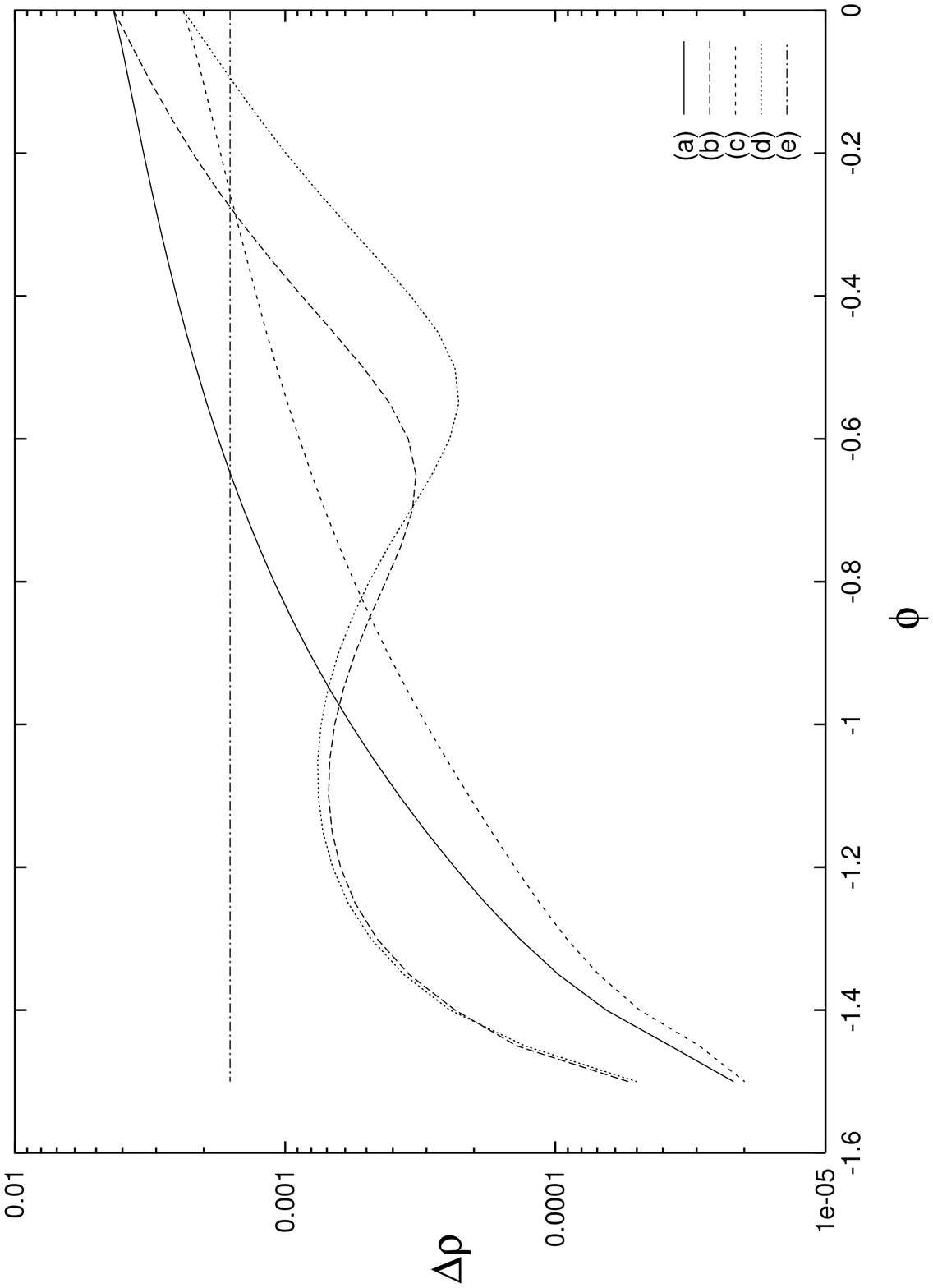,width=0.75\textwidth}}
\ccaption{}{ \label{fig:rho}
Contributions to the $\rho$ parameter from $\tilde{t}-\tilde{b}$
loops, for the cases (a) $k=1$ and tan$\beta =1$, (b) $k=1$ and tan$\beta
=5$, (c) $k=2$ and tan$\beta =1$, (d) $k=2$ and tan$\beta =5$,
compared with (e) the 95 \% C.L. after subtracting the Standard Model
prediction for $m_H = 100$ GeV.}

\end{figure}

We now
calculate ${\cal B}({\tilde t} \rightarrow e^+ s)$ in the
context of the ``strange stop" mechanism
$e^+ s_R \rightarrow {\tilde t}$ studied in this paper,
including stop
mixing effects~\footnote{In~\cite{AEGLM}, the corresponding calculation
of ${\cal B}({\tilde t}_L \rightarrow e^+ d)$ was presented, 
neglecting stop mixing and under the
alternative assumption that the ${\tilde t}_L$ was produced off
valence $d$ quarks in the proton, with a much smaller value
of the corresponding $\lambda'_{131}$ coupling.}.
In the presence of mixing, the $R$-violating 
${\tilde t}_1 \rightarrow e^+ s$ decay rate
becomes
\beq
\Gamma_{\tilde{t}_1} = \frac{1}{16 \pi} 
\lambda'^2_{132} m_{\tilde{q}}{\rm cos}^2{\phi}
\label{Rviolatingdecay}
\eeq
and the decay rate of $\tilde{t}_2$ is 
described by a similar equation, with cos$\phi$
substituted by sin$\phi$~\footnote{This decay would be
relevant if  $\tilde{t}_2$ were light enough to
have been produced at HERA, as proposed in \cite{KK}.}.
To illustrate our analysis, we choose values of the mixing angle $\phi$
that are consistent
with the $\Delta \rho$ bound shown in Fig.~2. 

The most important $R$-conserving decay is 
presumably that into a
chargino: ${\tilde t} \rightarrow \chi^+ b$. 
To understand this, we first note that
the decay ${\tilde t} \rightarrow \chi t$, where $\chi$
is the lightest neutralino, is
presumably excluded by LEP~2 lower limits on $m_{\chi}$, 
though a thorough analysis
in the appropriate $R$-violating framework has not yet been published.
Secondly, the kinematically-allowed ${\tilde t} \rightarrow \chi c$
decay is suppressed by typical loop factors, and three-body decays
are also expected to be small~\cite{AEGLM}. Since the
value of the $\lambda'_{132}$ coupling required in the
``strange stop" interpretation studied here ($\gappeq 0.3$) is much
larger than that of the corresponding $\lambda'_{131}$ coupling in the
``down stop" interpretation ($\gappeq 0.04$) 
pursued in~\cite{AEGLM}, 
it is easier for the $R$-violating decay mode to become competitive
with the $R$-conserving chargino decay mode, even when the latter is
not kinematically suppressed.

In the presence of mixing,
the ${\tilde t} \rightarrow \chi^+ b$
decay rate becomes~\cite{pioneers2}
\begin{eqnarray}
\Gamma' _{\tilde{t}_1} &&=
\frac{\alpha}{4sin^2\theta_W} \, m_{\tilde{t}_1} \,
\lambda^{1/2} (1, \frac{m_b^2}{m_{\tilde{t}_1}^2}
, \frac{m_{\chi^{\pm}}^2}{m_{\tilde{t}_1}^2} ) \,
\nonumber \\ &&  \cdot
\left\{\left(|G_L|^2+|G_R|^2\right)
\left(1- \frac{m_b^2}{m_{\tilde{t}_1}^2}-
\frac{m^2_{\tilde{\chi}^{\pm}}}{m_{\tilde{t}_1}^2}\right)
-{4m_b m_{\tilde{\chi}^{\pm}} \over m_{\tilde t_1}^2} {\rm Re}\left(G_R
G_L^*\right)\right\}, \\
 && G_L\equiv-{\frac{m_b U^*_{k2}{\rm cos}\phi}{\sqrt{2}m_W
{\rm cos}\beta}}, \,
\; \;  G_R\equiv V_{k1}{\rm cos}\phi+
{\frac{m_t V_{k2}{\rm sin}\phi}{\sqrt{2}m_W {\rm sin}\beta}}
\label{charginodecay}
\end{eqnarray}
where $\lambda(x,y,z)$ $\equiv$ $x^2+y^2+z^2-2xy-2yz-2zx$ is the usual
phase space factor, and the 
$V_{kl}$ and $U_{kl}$ 
are the chargino mixing angles, determined in the usual way by
$\mu$, the ratio tan$\beta$ of supersymmetric
Higgs vacuum expectation values, and $M_2$.
As compared to zero mixing,
we see that the vertex involving $\tilde{t}_L$
changes by a factor cos${\phi}$ like the
$R$-violating coupling. However, now
we also have the vertex that couples the chargino
to $\tilde{t}_R$, which is dependent on
$V_{1,2}$, $m_{t}$, and sin$\phi$.
Note also that
the branching ratios of the stop decays to charginos
depend on the sign of $\phi$, which is sensitive in particular
to the soft supersymmetry-breaking parameter $A$
and the Higgs mixing parameter $\mu$.

The presence of the additional vertex
suggests at first sight that
large stop mixing  might tend to favour the
$R$-conserving decay mode relative to the $R$-violating one.
Indeed, this feature is manifest in
the region of parameter space where
$\mu$ is small.
However, there is also the possibility of a
cancellation that may amplify the
$R$-violating branching ratio  in some specific regions
of parameter space. To see this, note that,
when stop mixing is taken into account, besides 
the squared couplings of 
the $\tilde{t}_{L,R}$  vertices, there also exists
a term in the decay rate arising from the cross product of
the parts of the 
$\tilde{t}_{R}$  and $\tilde{t}_{L}$  vertices
that have the same helicity~\footnote{The additional term
involving ${\rm Re}\left(G_R G_L^*\right)$
gives a much smaller contribution,
except in the small region with  
${m_{\tilde{t}_1}^2}-m^2_{\tilde{\chi}^{\pm}}
\sim m_b^2$, on which we do not comment
further.}. This ``cross term'' is proportional to
\beq
2 \; {\rm cos}\phi \; {\rm sin}\phi \; 
{\frac{m_t V_{k2} V_{k1}}
{\sqrt{2}m_W {\rm sin}\beta}}
\label{lab}
\eeq
This is significant,
and can be destructive if either
(i) sin$\phi$ is negative and $V_{11}$ and $V_{12}$ have the 
same sign, or
(ii) sin$\phi$ is positive and $V_{11}$ and $V_{12}$ have 
opposite signs.

It is clear that the stop branching ratio ${\cal B}(e^+ q)$
must be non-negligible, if H1 and ZEUS are to have seen this
decay mode without being drowned by other stop decay signatures.
On the other hand,
the D0~\cite{D0} and CDF~\cite{CDF} collaborations at the Fermilab ${\bar
p} p$ collider
have both established upper limits on $\sigma_{LQ} {\cal B}(e^+ q)^2$,
where $LQ$ is a generic scalar particle with decays into $e^+ q$,
which may for our purposes be taken as the $\tilde t_1$.
In particular, CDF~\cite{CDF} establishes that $m_{LQ} > 210$ GeV if
${\cal B}(e^+ q)
= 1$, corresponding to ${\cal B}(e^+ q) \lappeq 1 / \sqrt{1 + 9 \hbox{ln} 
\left ( \hbox{210 GeV} / {\mbox{$m_{LQ}$}} \right )}$
for values of the leptoquark mass below 210 GeV. 
The CDF limit~\cite{CDF} corresponds, in
particular, to ${\cal B} (e^+ q) < 0.73 (0.87)$ for $m_{LQ} = 190 (200)$
GeV, but provides no constraint for $m_{LQ} \ge 210$ GeV.

We combine the above constraints in the $\mu, M_2$ plane
as shown in Fig.~3, using the following
procedure. We first choose values of tan$\beta$ and of $\phi$,
the latter compatible with the
$\Delta \rho$ upper bound shown in Fig.~2. We then use
(\ref{Rviolatingdecay}) and (\ref{charginodecay}) to calculate 
${\cal B}(e^+ s)$, employing self-consistently the value of
$\lambda'_{132}$
suggested by the HERA observations: $\lambda'_{132} = 0.3/($cos$\phi
\sqrt{\cal B})$. We then apply the $Z \rightarrow e^+ e^-$ constraint of
Fig.~1 - which excludes models with low $\cal B$, 
in particular at small $M_2$ and/or $|\mu|$, and the CDF
constraint~\cite{CDF} - which excludes models with large $\cal B$ if
$m_{\tilde t_1} < 210$ GeV. As we see in Figs.~3(a,b), only restricted
regions of the $\mu, M_2$ plane are compatible with these constraints
if $m_{\tilde t_1} = 190$ GeV, and these are not greatly expanded if
one chooses $m_{\tilde t_1} = 200$ GeV as in Figs.~3(c,d). 
On the other hand, it is clear that the entire upper part of
the $\mu, M_2$ plane is allowed if $m_{\tilde t_1} \ge 210$ GeV,
since the CDF constraint~\cite{CDF} then imposes no restriction. 
Comparing
Figs.~3(a,c) with Figs.~3(b,d), we see that 
the allowed regions are 
of similar size, but more symmetric between positive and negative
$\mu$, if one chooses tan$\beta = 5$.

\begin{figure*}
\centerline{\epsfig{figure=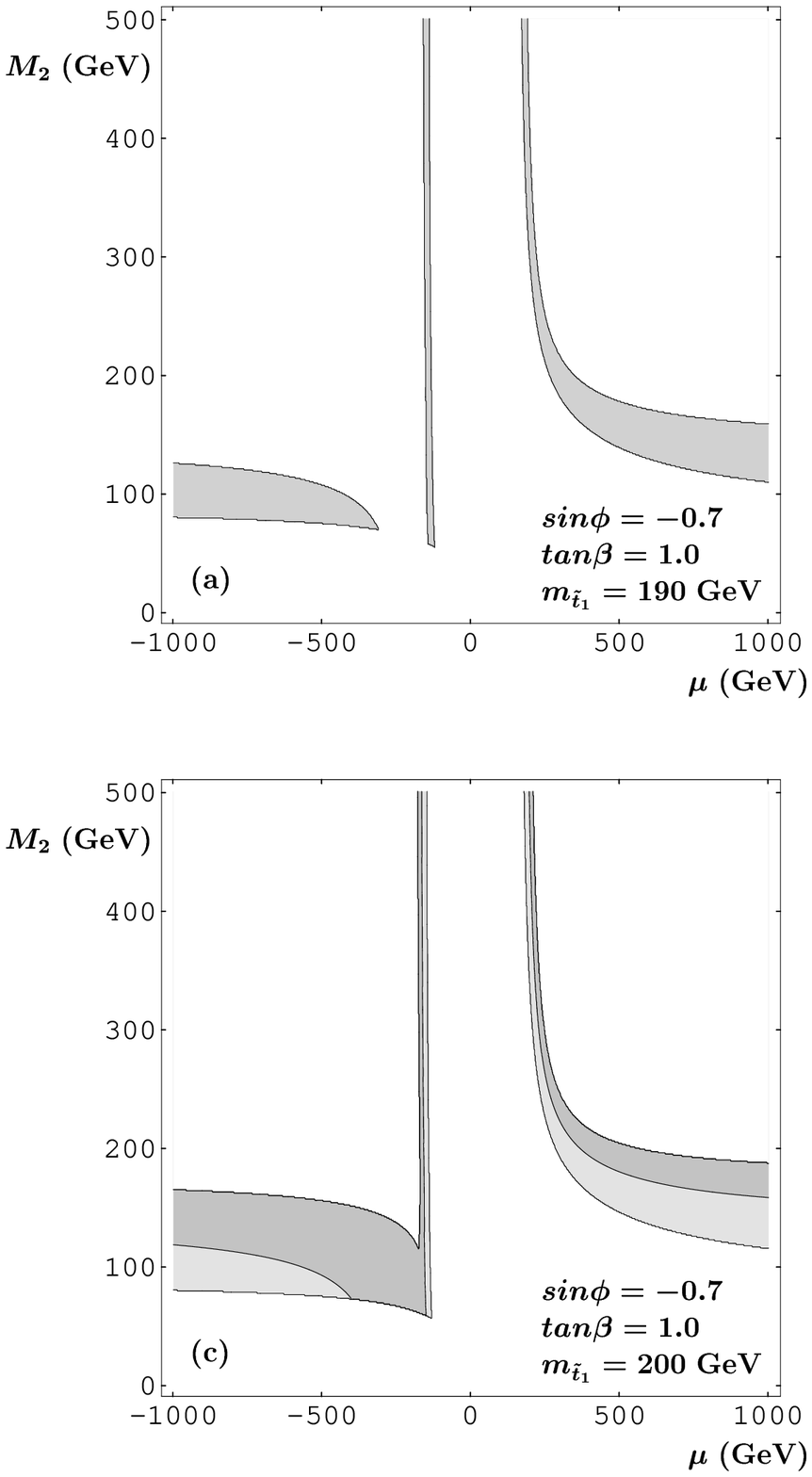,width=0.8\textwidth}
\hspace{-6 cm}\epsfig{
figure=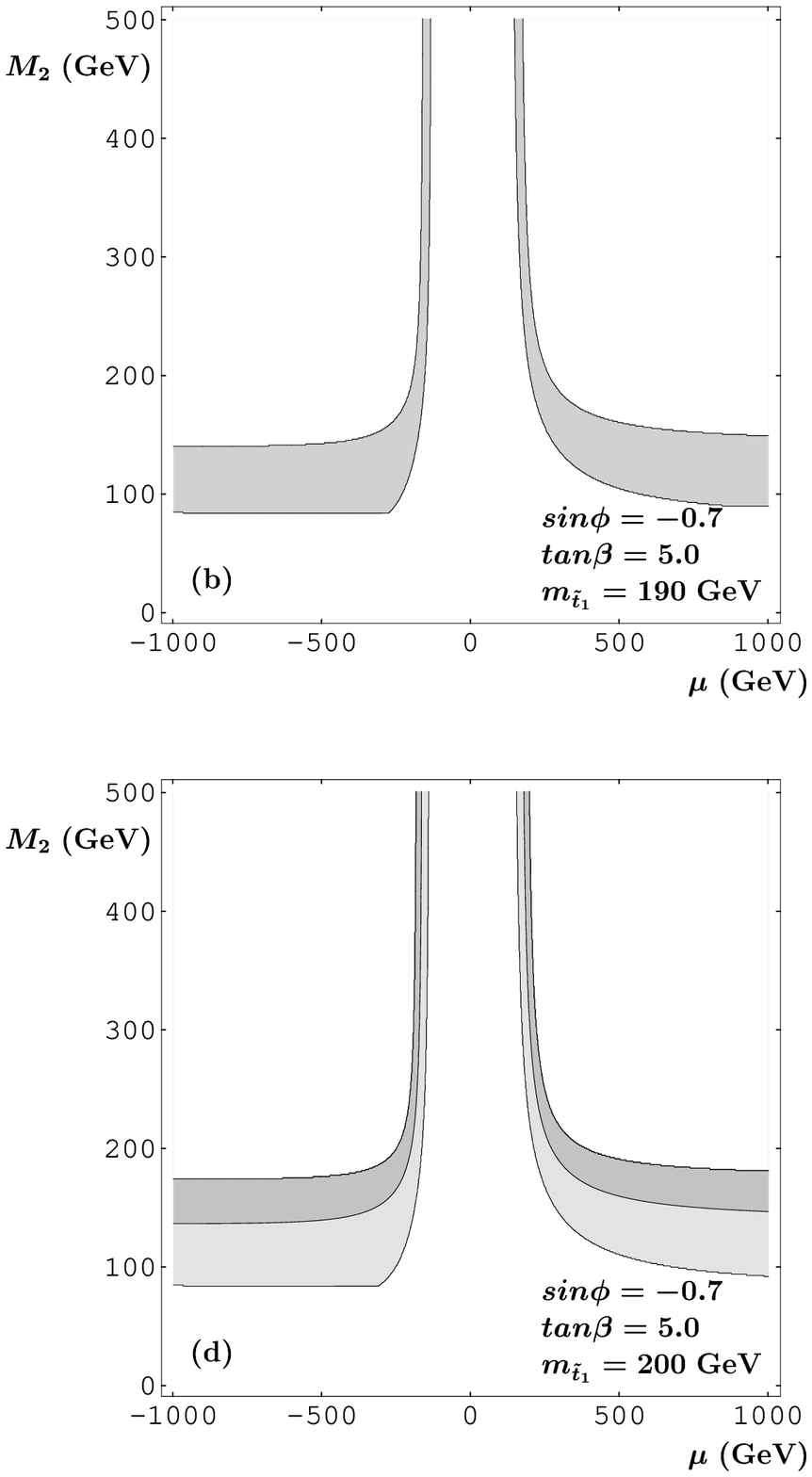,width=0.8\textwidth} }
\vspace*{-2 cm}
\caption{Regions of the $\mu,M_2$ plane that are consistent with
the upper bound on $\lambda'_{132}$ shown in Fig.~1 and the
CDF upper limit on ${\cal B}(e^+ q)$~\cite{CDF} for the
$R$-violating decay of $\tilde{t}_1$, for (a) sin$\phi = -0.7$,
tan$\beta = 1.0$ and $m_{\tilde t_1} = 190$ GeV, 
(b) sin$\phi = - 0.7$,
tan$\beta$=5.0 and the same value of
$m_{\tilde t_1}$, (c) sin$\phi = -0.7$,
tan$\beta = 1.0$ and $m_{\tilde t_1} = 200$ GeV, and (d) 
sin$\phi = - 0.7$, tan$\beta$=5.0 and the same value
of $m_{\tilde t_1}$.}
\end{figure*}

As was pointed out in~\cite{AEGLM}, one of the potentially
interesting constraints on leptoquark or $R$-violating
squark interpretations of the large-$Q^2$ HERA anomaly
could come from LEP~2 limits on effective contributions
to $e^+ e^- \rightarrow {\bar q} q$ due to crossed-channel
exchanges. Formulae have been given in~\cite{AEGLM} and
elsewhere for the case of single leptoquark or $R$-violating
squark exchange. However, the LEP~2 constraint is
potentially most relevant for the ``strange stop" interpretation
with its relatively large coupling: $\lambda'_{132} \sim 0.3 / 
(\hbox{cos} \phi \sqrt{\cal B})$, in which case stop mixing should be
taken 
into account in estimating the exchange contributions.
Accordingly, we now present for completeness the simple
generalization of the standard formulae 
available previously to the case in which
two mixed stops ${\tilde t}_{1,2}$ are exchanged:
\begin{eqnarray}
\sigma & =  & \sigma_{SM} + \frac{ 3 \lambda'^4 (I_1
{\rm cos}^4\phi+I_2 {\rm sin}^4\phi)} {64 \pi s} +
\frac{ 3 \lambda'^4 (I_3
{\rm cos}^2\phi {\rm sin}^2\phi)} {64 \pi s} +
\nonumber \\
& & + \frac{ 3 \lambda'^2 \alpha_{em}
(I'_1 {\rm cos}^2\phi+I'_2 {\rm sin}^2\phi)
}{4 s} \left[
e_e e_d +
a_L^e a_{R}^d             
\frac{s(s-M_Z^2)}{   
(s-M_Z^2)^2 + \Gamma_Z^2 M_Z^2}  \right ]
\label{dev}
\end{eqnarray}
where
\begin{eqnarray}
I_{1,2} & = & \frac{1+2x_{\tilde{t}_{1,2}}}{1+x_{\tilde{t}_{1,2}}} 
    -2x_{\tilde{t}_{1,2}}\ln\left( 
\frac{1+x_{\tilde{t}_{1,2}}}{x_{\tilde{t}_{1,2}}}\right)
\\  
I_3 & = & 
1 +\frac{x_{\tilde{t}_1}^2}{-x_{\tilde{t}_1}+x_{\tilde{t}_2}} 
\ln\left(\frac{1+x_{\tilde{t}_1}}{x_{\tilde{t}_1}}\right) 
- \frac{x_{\tilde{t}_2}^2}{-x_{\tilde{t}_1}+x_{\tilde{t}_2}} 
\ln\left(\frac{1+x_{\tilde{t}_2}}{x_{\tilde{t}_2}}\right) \\
I'_{1,2} & = & -\frac{1}{2}+x_{\tilde{t}_{1,2}}-x_{\tilde{t}_{1,2}}^2\ln 
    \left( \frac{1+x_{\tilde{t}_{1,2}}}{x_{\tilde{t}_{1,2}}}\right) 
\end{eqnarray}
with  
$a_{L,R}^f = (T_3^f - e^f \sin^2\theta_W)/ (\sin\theta_W
\cos\theta_W)$ and 
$x_{\tilde{t}_{1,2}} \equiv \frac{m^2_{\tilde{t}_{1,2}}}{s}$. 
The contributions 
proportional to $I_{1,2}$ in eq.(\ref{dev}) 
arise from the squared amplitudes of the
diagrams with the $R$-violating     
vertices, for $\tilde{t}_1$  and $\tilde{t}_2$  
exchange respectively, and $I_3$ is the interference term
between the  $\tilde{t}_1$  and $\tilde{t}_2$  exchange diagrams.
Finally, the terms proportional to $I'_{1,2}$ are 
interference terms with
the Standard Model $s$-channel $\gamma$ and $Z$ exchanges,
for both stop exchanges~\footnote{Note that, as well as the different
coupling factors, because of fermionic Wick ordering
there is a difference in the sign of the interference term
for charge-1/3 ${\bar q} q$ final states such as the ${\bar s} s$
case considered here, 
as compared to charge-2/3 final states such as the ${\bar c} c$
case considered in~\cite{AEGLM}. This has the consequence that the
interference terms are numerically negative in both cases, as shown
in~\cite{KRSZ}: we thank P.~Zerwas for discussions on this point.}.

The interference of the new diagrams with the
Standard Model processes is in our case destructive.
In the case we discussed previously~\cite{AEGLM} with $\lambda'_{131} =
0.04$, interference gave the dominant
deviation from the Standard Model prediction. However, in the
case studied in this paper with $\lambda'_{132} \lappeq 0.3$, 
the squares of the new diagrams are also large, and cancellations
occur between them and the negative interference terms.
Moreover, when mixing is introduced, the squared terms and the
destructive interference ones are multiplied by 
different factors,
in such a way that the cancellation is {\it enhanced}.
Thus the bounds derivable from LEP~2 bounds may be
relaxed significantly compared to the unmixed case.
Indeed, for any given value of $\lambda'_{132}$, there is even a value of
$\phi$ where no bound is obtained, since
the cancellation becomes exact. More generally, when
two light stops are exchanged, the cancellation 
occurs over a wider range of
mixing angle $\phi$, and the LEP~2 bounds are relaxed.
The numbers for some representative parameter choices
are shown in the Table. For various different values of
$m_{\tilde t_2}$ and $\phi$, the magnitudes of the
destructive interference terms and the new squared
amplitude contributions are shown, and we see how they
conspire to leave small net contributions to the cross section for
$e^+ e^- \rightarrow {\bar s} s$.

\begin{table}[t]
\centering
\begin{tabular}
{|c|c|c|c|}
\hline
\multicolumn{4}{|c|}
{ } \\ 
\multicolumn{4}{|c|}
{$\lambda'_{132} = 0.4 $, \,
$m_{\tilde{t}_1}$ = 200 GeV, \,
$m_{\tilde{t}_2}$ very heavy} \\
\multicolumn{4}{|c|}
{ } \\ 
\hline 
$|sin\phi|$   & 
cross section (pb)   & 
squared amplitude (pb) &
negative interference(pb) \\
\hline \hline
 0.0 & 0.082 & 0.415 & -0.333 \\
 0.2 & 0.063 & 0.383 & -0.320 \\
 0.4 & 0.013 & 0.293 & -0.280 \\
 0.6 & -0.043 & 0.170 & -0.213 \\
 0.8 & -0.066 & 0.054 & -0.120 \\
\hline
\end{tabular}
\begin{tabular}
{|c|c|c|c|}
\multicolumn{4}{c}
{ } \\ 
\hline
\multicolumn{4}{|c|}
{ } \\ 
\multicolumn{4}{|c|}
{$\lambda'_{132} = 0.4 $, \,
$m_{\tilde{t}_1}$ = 200 GeV, \,
$m_{\tilde{t}_2}$ = 205 GeV} \\
\multicolumn{4}{|c|}
{ } \\ 
\hline 
$|sin\phi|$   & 
cross section (pb)   & 
squared amplitude (pb) &
negative interference (pb) \\
\hline \hline
 0.0 & 0.082 & 0.415  & -0.333 \\
 0.2 & 0.066 & 0.399  & -0.333 \\
 0.4 & 0.025 & 0.356  & -0.331 \\
 0.6 & -0.016 & 0.313 & -0.329 \\
 0.8 & -0.020 & 0.306 & -0.326 \\
\hline
\end{tabular}
\caption{
\small{The contribution additional to the Standard Model
cross section for
$e^+ e^- \rightarrow s\bar{s}$ at $E_{CM} = 192$ GeV, for different values
of the stop mixing angle $\phi$. 
The second column gives the net
additional contribution to the
cross section due to the new terms. The third column
presents the positive contributions due to the
new squared amplitude terms in the first line of (\ref{dev}),
and the fourth column presents the negative terms due to the
destructive interference in the second line of (\ref{dev}).}}
\label{table:1}
\end{table}

We have seen in the above analysis that the ``strange stop"
scenario may survive the available constraints, at the price of
requiring a relatively restricted region in the $\mu, M_2$ plane
if $m_{\tilde t_1} \le 200$ GeV. It is disappointing that
partial cancellations between negative interference and positive
squared-amplitude terms may hinder attempts to pin down this
scenario at LEP~2. In principle, the decay modes and branching
ratios observable at HERA could discriminate between the ``down" and
``strange stop" scenarios. There are also
possibilities for discriminating between these scenarios by
using different beams at HERA.

One clear distinction could be drawn by comparing stop production
in $e^+ p$ and $e^- p$ collisions. In the ``down stop" case, one
predicts $\sigma (e^+ p) >> \sigma(e^- p)$, since production occurs
off valence quarks in $e^+ p$ collisions, and off sea quarks in
$e^- p$ collisions. One does not expect such a large ratio in the
``strange stop" interpretation, since production occurs off sea quarks
with either beam. However, it is not necessarily the case that the
production cross sections must be exactly equal, since the $s$ and $\bar
s$ distributions may have different shapes, though their integrals must be
equal. If the strange sea in the proton is extrinsic, e.g.,
generated by perturbative QCD, one would expect
strict equality in the cross sections, but this need not be the
case if there is an intrinsic~\cite{chevy} strange
component~\footnote{There is no
indication of this in the data available so far, but we do
note that there are indications of other
flavour asymmetries in the ${\bar q} q$ sea~\cite{gott}.}. Thus, we
predict only
that $\sigma (e^+ p) \sim \sigma (e^- p)$ in the ``strange stop"
interpretation.

Another tool for discriminating between models could be provided
by polarized beams at HERA. Both the ``down stop" and ``strange stop"
mechanisms involve the component of the $e^+$ beam with right-handed
longitudinal polarization. This means that dramatic effects should be 
seen once longitudinal {\it lepton} beam polarization is commissioned at
HERA: both
models predict that $\sigma (e^+_L p) <<< \sigma (e^+_R p)$ and
$\sigma (e^-_R p) <<< \sigma (e^-_L p)$.
Polarized {\it proton} beams~\cite{polprot} would also be very
interesting. In the case
of ``down stop" model, since most models predict that valence down quarks
should mainly be polarized {\it opposite} to the proton spin, one would
expect that $\sigma (e^+_R p_R) << \sigma (e^+_R p_L)$. 

The predictions for polarized proton beams in
the ``strange stop" model are currently less clear, since there is
no consensus on the amount of polarization $\Delta s$ of the ${\bar s} s$
sea
in the proton. The most straightforward interpretation of the available
data on polarized deep-inelastic $\ell N$ scattering is that $\Delta s <
0$~\cite{dels}, though this suggestion is subject to possible gluonic
reinterpretation~\cite{delg}. However, the available data give no clear
indication
on the {\it degree} of polarization $\Delta s / ({\bar s} + s)$,
particularly not
at the large $x$ values probed by the HERA experiments, nor do we
know how $\Delta s$ might be shared between the $\bar s$ and $s$ in the
proton. One particularly naive model suggests that the $s$ quarks may
have a large degree of polarization, particularly at large $x$, and
that the same {\it might} also be true of the $\bar s$~\cite{ekks}. In
this case, one
might expect that
\begin{equation}
\sigma (e^+_R p_R) < \sigma (e^+_R p_L) \, , \,
\sigma (e^-_L p_L) < (?) \sigma (e^-_L p_R)
\label{bothpol} 
\end{equation}
with much smaller cross sections for the other beam
polarizations, as mentioned above.

It is very likely that the mystery of the large-$Q^2$ HERA
anomaly will be resolved by the time such polarized experiments
become possible. However, these examples serve to illustrate
that both beam polarizations could be useful analysis tools.

\vspace*{0.2 cm}
{\it Acknowledgements:} 
We would like to thank Michelangelo Mangano for discussions.
One of us (K.S.) would like to thank
the Theory Division, CERN and the Centre for Particle Theory,
Durham for hospitality during the period this work was completed.
K.S. would also like to thank the PPARC, U.K. for a Fellowship
during his stay in Durham. The work of S.L. is funded by a
Marie Curie Fellowship (TMR-ERBFMBICT-950565).

\end{document}